\documentclass{article}

\def\no{\, : \,}

\def\kp{\,\dot{+}\,}

\begin{document}

\title{%
      Lorentz invariance of scalar field action on $\kappa$-Minkowski space-time}
\author{ Sebastian {Nowak}\thanks{e-mail: {\tt pantera@ift.uni.wroc.pl}}\\  \\ {\em Institute for Theoretical
Physics}\\ {\em University of Wroc\l{}aw}\\ {\em Pl.\ Maxa Borna 9}\\{\em Pl--50-204 Wroc\l{}aw, Poland}} \maketitle

\begin{abstract}
We construct field theory on noncommutative $\kappa$-Minkowski
space-time. Having the Lorentz action on the noncommutative
space-time coordinates we show that the field lagrangian is
invariant. We show that noncommutativity requires replacing the
Leibnitz rule with the coproduct one.
\end{abstract}

\section{Introduction}
$\kappa$-Minkowski space-time with no-trivial commutator between time and space being
\begin{equation}\label{1}
  [x_0, x_i] = - \frac i\kappa \, x_i
\end{equation}
has been first constructed in \cite{zakrz}, \cite{MajidPLB334} as
a Hopf algebra dual to the translational part of
$\kappa$-Poincar\'e algebra \cite{qP}. After Doubly Special
Relativity (DSR) has been formulated in
\cite{Amelino-Camelia:2000ge}, \cite{jkgminl}, \cite{rbgacjkg}, it
was soon realized \cite{Kowalski-Glikman:2002jr} that
$\kappa$-Minkowski space-time is a natural candidate for
space-time of DSR\footnote{For up to date reviews of DSR program
see \cite{Kowalski-Glikman:2003hi},
\cite{Amelino-Camelia:2002vy}.}. This space-time arises also
naturally in investigations of 2+1 (quantum) gravity
\cite{Matschull:1997du}, \cite{Freidel:2003sp}, giving rise to the
claim that it is the space-time emerging from semiclassical, weak
field approximation of 3+1 gravity as well.

It seems natural to investigate properties of classical and
quantum fields on $\kappa$-Minkowski space-time. Indeed such
investigations has been undertaken, among others in
\cite{Kosinski:1998kw}, \cite{Kosinski:1999ix},
\cite{Kosinski:2001ii}, \cite{Amelino-Camelia:2001fd},
\cite{Kosinski:2003xx}, \cite{Dimitrijevic:2004nv}. It has been
shown \cite{GAChopf} that the invariance of action on scalar
fields on $\kappa$-Minkowski leads to deformed algebra of
symmetries and some nontrivial co-algebra structure.  In this
paper we proceed in different manner, we consider Lorentz action
on generators of $\kappa$-Minkowski space and compute Lorentz
action on normally ordered plane wave  and then we show Lorentz
invariance of $\kappa$-deformed Klein-Gordon action.

\section{$\kappa$-deformed algebra of symmetries}

In this section we review the construction of $\kappa$-Poincar\'e
algebra \cite{qP} which can be seen as deformed algebra of
symmetries. Lorentz sector of this algebra is standard but the
action of boost generators on momenta is deformed. Here we write
down this algebra in so-called bicrossproduct  basis
\cite{MajidPLB334}
$$
[M_i, M_j] = i\, \epsilon_{ijk} M_k, \quad [M_i, N_j] = i\,
\epsilon_{ijk} N_k,
$$
\begin{equation}\label{1}
  [N_i, N_j] = -i\, \epsilon_{ijk} M_k.
\end{equation}
$$
  [M_i, p_j] = i\, \epsilon_{ijk} p_k, \quad [M_i, p_{0}] =0
$$
\begin{equation}\label{2}
   \left[N_{i}, {p}_{j}\right] = i\,  \delta_{ij}
 \left( {\kappa\over 2} \left(
 1 -e^{-2p_{0}/\kappa}
\right) + {{\mathbf{p}^2}\over 2\kappa}  \right) -
\frac{i}{\kappa} p_{i}p_{j} ,\,\,\, \left[N_{i},p_{0}\right] = i\,
p_{i}.
\end{equation}
This algebra together with coalgebra structure written below forms noncommutative, noncocommutative Hopf algebra
$$
\triangle (M_i) =M_i\otimes 1 + 1\otimes M_i
$$
$$
 \triangle (N_i)=N_i \otimes 1 +e^{-p_0/\kappa}\otimes N_i +\frac{1}{\kappa}\epsilon_{ijk}p_j \otimes M_k
$$
\begin{equation}\label{3}
 S(N_i)=-e^{p_0/\kappa}(N_i-\frac{1}{\kappa}\epsilon_{ijk}p_j M_k ),\,\,\, S(M_i)=-M_i
\end{equation}
$$
 \triangle (p_i)=p_i\otimes 1 +e^{-p_0/\kappa}\otimes p_i
$$
$$
\triangle (p_0)=p_0 \otimes 1 +1\otimes p_0
$$
\begin{equation}\label{4}
   S(p_i)=-p_ie^{p_0/\kappa} ,\,\,\, S(p_0)=-p_0
\end{equation}
Following the general scheme we can introduce $\kappa$-Minkowski
space $T^*$ as a dual Hopf algebra to algebra of translations
(momenta). With the use of  pairing
\begin{equation}\label{60}
    <p_{\mu},x_{\nu}>=-i\eta_{\mu\nu}
\end{equation}
together with axioms of a Hopf algebra duality
$$
<t,xy>=<t_{(1)},x><t_{(2)},y>,
$$
$$
<ts,x>=<t,x_{(1)}><s,x_{(2)}>,
$$
\begin{equation}\label{61}
  \forall t,s \in T,\;\;\;\; x,y \in T^{*}
\end{equation}
we get
$$
[x_{i},x_{j}]=0,\;\;\;\; [x_{0},x_{i}]=-\frac{i}{\kappa}x_{i}
$$
\begin{equation}\label{62}
    \triangle x_{\mu}=x_{\mu}\otimes 1+1\otimes x_{\mu}.
\end{equation}
where we use Sweedler notation for coproduct
$$
\triangle t=\sum t_{(1)}\otimes t_{(2)}.
$$
Having the Hopf algebra of space-time coordinates we can write down the covariant action of T on it:
\begin{equation}\label{63}
    t\triangleright x=<x_{(1)},t>x_{(2)}, \;\;\; \forall x\in T^{*}, \;\;\; t\in T.
\end{equation}
In our case it reads
$$
p_{i}\triangleright :\psi
(x_{i},x_{0}):=\frac{1}{i}:\frac{\partial}{\partial x_{i}} \psi
(x_{i},x_{0}):,
$$
\begin{equation}\label{64}
    p_{0}\triangleright :\psi (x_{i},x_{0}):=-\frac{1}{i}:\frac{\partial}{\partial x_{0}} \psi (x_{i},x_{0}):
\end{equation}
which is  like in classical case but remembering normal ordering.

 The  action of U(so(1,3)) generators on momenta, defined in formula (\ref{2}) (we denote it here for simplicity by
 $\triangleleft$)  , can be translated by duality  into generators of $\kappa$-Minkowski space. We have
$$
<t,h\triangleright x>=<t\triangleleft h,x>
$$
\begin{equation}\label{66}
    \forall t \in T,\;\;\; h\in U(so(1,3)),\;\;\; x\in T^{*}.
\end{equation}
 Using duality pairing written in (\ref{60}) we get
$$
 M_{i}\triangleright x_{0}=0,\;\;\; M_{i}\triangleright x_{j}=i\epsilon_{ijk}x_{k},
$$
 \begin{equation}\label{67}
 N_{i}\triangleright x_{0}=ix_{i},\;\;\; N_{i}\triangleright x_{j}=i\delta_{ij}x_{0}.
\end{equation}
We see that the action is the same as in classical case. The
difference occurs while acting on product of coordinates. In
noncommutative case we must use the coproduct formula instead of
usual Leibnitz rule so as no contradiction with relation
$$
[x_{0},x_{i}]=-\frac{i}{\kappa}x_{i}
$$
arises. We have
\begin{equation}\label{68}
    h\triangleright (xy)=(h_{(1)}\triangleright x)(h_{(2)}\triangleright y)
\end{equation}
Moreover from the structure of bicrossproduct construction it appears that one can also define the action of
space-time algebra on Lorentz algebra and then define the cross relations
between them \cite{MajidPLB334}. In our case it has following form
$$
[M_{i},x_{0}]=0,\;\;\; [M_{i},x_{j}]=i\epsilon_{ijk}x_{k},
$$
\begin{equation}\label{A}
  [N_{i},x_{0}]=ix_{i}-\frac{i}{\kappa}N_{i},\;\;\; [N_{i},x_{j}]=i\delta_{ij}x_{0}-\frac{i}{\kappa}\epsilon_{ijk}M_{k}.
\end{equation}
It has been shown that so(1,3) algebra and space-time algebra (\ref{61}) together with
above cross relations form so(1,4) algebra. Moreover the action on momenta written in equation
(\ref{2}) may be derived from the action of so(1,4) algebra on four dimensional de Sitter space of momenta \cite{deSitter}.

\section{Integration and Fourier transform}

In order to define Fourier transform on $\kappa$-Minkowski we have
to define normally ordered plane wave, which is a solution of
$\kappa$-deformed Klein-Gordon field equation
\cite{Kosinski:2003xx}.
 Normal ordering is necessary because  space-time coordinates
do not commute, here we choose  "time to the right" ordering
\begin{equation}\label{B1}
    :e^{ipx}:=e^{i\mathbf{p}\mathbf{x}}e^{-ip_0x_0}
\end{equation}
 which simply means all $x_0$ shifted to the right.
Having normally ordered plane wave one can define $\kappa$-deformed Fourier transform describing fields on
noncommutative space:
\begin{equation}\label{74}
    \Phi(x_0, \mathbf{x})=\frac{1}{(2\pi)^{4}}\int d\mu \Phi(p_0, \mathbf{p})\,e^{i\mathbf{p}\mathbf{x}}\,e^{-ip_0 x_0} ,
\end{equation}
where
\begin{equation}\label{75}
    d\mu =d^{4}p\, e^{3p_0/\kappa}
\end{equation}
is the measure invariant under the action of U(so(1,3)) algebra.
The field written  above is  an element of Hopf algebra described
in equations (\ref{62}). One can equip this algebra with hermitian
conjugation. This implies the conjugation of field:
\begin{equation}\label{F1}
  \Phi^{+}(x_0, \mathbf{x})=\frac{1}{(2\pi)^{4}}\int d\mu \Phi^{\star}(p_0, \mathbf{p})\,e^{ip_0 x_0}\,e^{-i\mathbf{p}\mathbf{x}} ,
\end{equation}
which can be rewritten in the following form:
\begin{equation}\label{F2}
     \Phi^{+}(x_0, \mathbf{x})=\frac{1}{(2\pi)^{4}}\int d\mu \Phi^{\star}(p_0, \mathbf{p})\,e^{iS(\mathbf{p})\mathbf{x}}\,e^{-iS(p_0) x_0} ,
\end{equation}
where we used equality
\begin{equation}\label{F4}
    e^{ip_0\,x_0}e^{-i\mathbf{p}\,\mathbf{x}}=e^{-i\mathbf{p}e^{p_0/\kappa}\,\mathbf{x}}e^{ip_0\,x_0}
\end{equation}
together with definition of antipode S (equation (\ref{4})).

In order to define the field theory on $\kappa$-Minkowski we need
to define $\kappa$-deformed integration. To end this we must
introduce, in analogy to the classical case, the following formula
\begin{equation}\label{76}
    \frac{1}{(2 \pi)^{4}}\int\int d^{4}x :e^{ipx}:=e^{3p_0/\kappa}\delta^{4}(p)
\end{equation}
This formula is invariant under the action of Lorentz generators
on momentum space defined in equations (\ref{2}) . Let us check it
for boost transformation; we have
\begin{equation}\label{77}
  \frac{1}{(2 \pi)^{4}}\int\int d^{4}x :e^{ip'x}:=e^{3p'_{0}/\kappa}\delta^4 (p')=e^{3p'_{0}/\kappa}J(\frac{\partial p'_{\mu}}{\partial p_{\nu}})\delta^4(p)=e^{3p_{0}/\kappa}\delta^4 (p)
\end{equation}
and the same result holds for rotation generators.

Using the above definition of delta function one can find
$$
\int \, d^4x \Psi(x) \Phi(x) =
$$
$$
\int  d^4x \int d\mu d\mu'\,   \,  \tilde{\Psi}(p)
\tilde{\Phi}(p')\no e^{ipx} \no  \no e^{ip'x} \no =
$$
\begin{equation}\label{25}
   \int\, d\mu\,  d\mu' \, \tilde{\Psi}(p)\, \tilde{\Phi}(p')\, \delta(p \kp p')
\end{equation}
To derive this formula one uses the fact that
\begin{equation}\label{26}
    \no e^{ipx} \no \no e^{ip'x} \no = \no e^{i(p \kp p')x} \no
\end{equation}
where
\begin{equation}\label{27}
    p \kp p' = (p_0+p'_0; \mathbf{p} +  e^{-p_0/\kappa} \,\mathbf{p'})
\end{equation}
represents the ``co-product summation rule'', related to the Hopf
algebra structure of $\kappa$-Poincare algebra.

We can define the scalar product of fields on $\kappa$-Minkowski
as
\begin{equation}\label{F3}
    \left(\Phi(x),\Psi(x)\right)=\int d^4x\,\Phi^+(x)\,\Psi(x)=\int d^4p\,e^{3p_{0}}\,\Phi^*(p)\Psi(p)
\end{equation}

Now it is a good point to introduce Lorentz action on fields on
noncommutative space-time. We start with normally ordered plane
wave. Using the Lorentz action on noncommuting coordinates
(\ref{67}) together with coproduct rule
 (\ref{68})  we find
$$
M_{i}\triangleright (e^{ip_{j}x_{j}}e^{-ip_{0}
x_{0}})=\epsilon_{ijk}x_{j}p_{k}e^{ip_{j}x_{j}}e^{-ip_{0}x_{0}}
$$
$$
N_{i}\triangleright (e^{ip_{j}x_{j}}e^{-ip_{0}x_{0}})=
$$
\begin{equation}\label{u1}
  e^{ip_{j}x_{j}}\left[  \left(\delta_{ij}(\frac{\kappa}{2}(1-e^{-\frac{2P_{0}}{\kappa}})
    +\frac{1}{2\kappa}\vec{p}^{\,2})-\frac{1}{\kappa}p_ip_j\right)x_j+
    (-p_ix_{0})\right]e^{-ip_{0}x_{0}}
\end{equation}
and together with action of translation generators defined in
equation (\ref{64}) we have the action of the $\kappa$-Poincar\'e
algebra on normal ordered plane wave. One can easily show that the
generators of this action are hermitian with respect to scalar
product (\ref{F3}).  One can also find differential realization of
the above action \cite{Kosinski:2003xx}, which turns out to be
nonlinear.

Now taking infinitesimal parameter $\varepsilon$ we can introduce infinitesimal Lorentz transformations on normal ordered
plane waves.
  $$
(1+i\varepsilon M_{i})\triangleright (e^{ip_{j}x_{j}}e^{-ip_{0}
x_{0}})=e^{ip'_{j}x_{j}}e^{-ip_{0}x_{0}}
$$
where
\begin{equation}\label{69}
p'_{j}=p_{j}-i\varepsilon  [M_{i},p_{j}]
\end{equation}
and
$$
(1+i\varepsilon N_{i})\triangleright
(e^{ip_{j}x_{j}}e^{-ip_{0}x_{0}})=
 e^{i p'_{j}x_{j}}e^{-ip_{0}'x_{0}}
$$
\begin{equation}\label{70}
    p'_{j}=p_{j}-i\varepsilon [N_{i},p_{j}],\;\;\; p'_{0}=p_{0}-i\varepsilon [N_{i},p_{0}]
\end{equation}

We see that our normally ordered plane wave is an analog of classical plane wave $e^{ipx}$.
The action on space-time coordinates interchanges with the action on energy-momentum space. Moreover in both cases one
 can see
the phase space as the Lee algebra of space-time coordinates and
the energy-momentum Hopf algebra as the algebra of functions on
group generated by the algebra of space-time coordinates. The
significant difference is that the classical algebra is
commutative while the quantum is not.

Knowing the action of Lorentz algebra on plane wave we can write
down the transformations of Fourier transform (\ref{74})
\begin{equation}\label{71}
   (1+i\varepsilon N_{i})\triangleright \Phi (x,t)=\int d\mu \Phi(p_{0},\vec{p}) e^{i p'_{j}x_{j}}e^{-ip_{0}'x_{0}}.
\end{equation}
Now we can change variables under the integral and  since the
measure $d \mu$ is invariant under the action of U(so(1,3))
algebra we get
\begin{equation}\label{72}
   (1+i\varepsilon N_{i})\triangleright \Phi (x,t)=\int d\mu \Phi(p'_{0},\vec{p'}) e^{i p_{j}x_{j}}e^{-ip_{0}x_{0}}
\end{equation}
where
\begin{equation}\label{73}
   p'_{j}=p_{j}+i\varepsilon [N_{i},p_{j}],\;\;\; p'_{0}=p_{0}+i\varepsilon [N_{i},p_{0}]
\end{equation}
but on the other side we can write
\begin{equation}\label{73a}
    (1+i\varepsilon N_{i})\triangleright \Phi (x,x_0)=\Phi'(x,x_0)=\int d\mu \Phi'(p_{0},\vec{p}) e^{i p_{j}x_{j}}e^{-ip_{0}x_{0}}.
\end{equation}
Compering equations (\ref{72}) and (\ref{73a}) we see that
$$
\Phi'(p'_{0},\vec{p'})=\Phi(p_{0},\vec{p})
$$
which means that fields in energy-momentum space are scalar
fields. Here we consider only the action of algebra but one can
also consider the action of $\kappa$-Poincar\'e group on fields.
This  was done in paper \cite{Kosinski:2001ii}, with the help of
$\kappa$-Wigner construction.

Using the first order Taylor expansion we can write  equation
(\ref{72}) in the following form
\begin{equation}\label{80}
 (1+i\varepsilon N_{i}) \triangleright \Phi (x,t)=\int d\mu (1+i\varepsilon N_{i} ) \Phi(p_{0},\vec{p}) e^{i p_{j}x_{j}}e^{-ip_{0}x_{0}},
\end{equation}
where the boost generator on the right hand side stands for differential operator in momentum space. Having the above
transformation rules we see that  variables $p_0,
p_i$ may be indeed identified with energy and momentum in
bicrossproduct basis.

Another important thing to notice is the action of
$\kappa$-Poincar\'e algebra on product of fields. To end this we
must apply the coalgebra structure, which in classical case, where
the coproducts are trivial, is just the Leibnitz rule. The action
of Lorentz algebra is due to the covariance condition (\ref{68}).
We get
$$
  N_i \triangleright (\Phi(x)\Psi(x))=\int\int d^4 p\,e^{3p_0/\kappa}\,d^4 k\,e^{3k_0/\kappa}\,
  \left[ ( N_{i}\Phi(p))\Psi(k)+\right.
$$
\begin{equation}\label{78}
 \left.  e^{-p_0/\kappa}
  \Phi(p)(N_{i} \Psi(k))+ \epsilon_{ijk}p_j \Phi(p)( M_{k} \Psi(k))\,\right]:e^{ipx}::e^{ikx}:.
\end{equation}
From the above equation it is seen that noncommutativity of
space-time enforces applying the coproduct rule on fields in
momentum space \cite{Daszkiewicz:2004xy}.  The action of
translation generators is due to the equation (\ref{64}) and it
reads
$$
p_0 \triangleright (\Phi(x)\Psi(x))=\int\int d^4
p\,e^{3p_0/\kappa}\,d^4 k\,e^{3k_0/\kappa}\,(p_0
+k_0)\Phi(p)\Psi(k):e^{ipx}::e^{ikx}:
$$
\begin{equation}\label{79}
    p_i \triangleright (\Phi(x)\Psi(x))=\int\int d^4 p\,e^{3p_0/\kappa}\,d^4 k\,e^{3k_0/\kappa}\,(p_i +e^{-p_0/\kappa}k_i)\Phi(p)\Psi(k):e^{ipx}::e^{ikx}:
\end{equation}
which is nothing but the coproduct summation rule. In the next
section we'll see that we can also define invariant action and
field equation.

\section{Actions and field equations}

  We consider here only the free $\kappa$-deformed KG (Klein-Gordon) action which has the following form \cite{Kosinski:1999ix}:
\begin{equation}\label{u2}
  S=\frac12\int d^4 x \,\left(\eta^{\mu\nu} \, \Phi(x)\, \hat{\partial}_\mu
  \hat{\partial}_\nu\, \Phi(x) - \mathcal{M}^2\, \Phi(x)^2\right).
\end{equation}
where $\hat{\partial}_\nu$ means covariant differentiation on
noncommuting space-time \cite{Woron}, \cite{Gon}, \cite{Sitarz}.
We consider only the real fields ($\Phi^+(x)=\Phi(x)$). By
construction the action should be invariant under action of
deformed $\kappa$-Poincar\'e algebra. Let us show   this
invariance explicitly. To do that we write the action in somewhat
convenient form
\begin{equation}\label{Q1}
    S=\int d^4 x \, \int \int d^{4}p\,e^{3p_0/\kappa}\,d^{4}k\,\Phi(p)\Phi(S(k)):e^{ipx}::e^{iS(k)x}:\mathcal{M}(k)
\end{equation}
where $\mathcal{M}(k)$ is the mass shell condition and  will be
explicitly written below. Using formula (\ref{78}) we have
$$
(1+i\varepsilon N_{i})\triangleright S=\int d^{4}x\,\int\int d^4
p\,e^{3p_0/\kappa}\,d^4 k\,
$$
$$
\left[ \Phi(p)\Phi(S(k))+(i\varepsilon N_{i}\Phi(p))\Phi(S(k))+
e^{-p_0/\kappa}\Phi(p)\left(-i\varepsilon S(N_{i})
\Phi(S(k))\,\right)\right.
$$
\begin{equation}\label{Q2}
 \left. +\epsilon_{ijk}p_j \Phi(p)\left(-i\varepsilon S(M_{k}) \Phi(S(k))\,\right)\,\right]\mathcal{M}(k):e^{ipx}::e^{iS(k)x}:.
\end{equation}
Now using the following formulas for antipodes S
\begin{equation}\label{Q3}
    S(M_i)=-M_i,\,\,\, S(N_i)=-e^{p_0/\kappa}(N_i -\frac{1}{\kappa}\epsilon_{ijk}p_j M_k )
\end{equation}
together with definition of delta function (\ref{76}) we get
\begin{equation}\label{Q4}
    (1+i\varepsilon N_{i})\triangleright S=\int d^{4}p\,e^{3p_{0}/\kappa}\,(1+\varepsilon N_i )\left(\Phi(p)\Phi(S(p))\,\right)\mathcal{M}(k)=S
\end{equation}
where we made use of the invariance of the measure. We get the
same result if we repeat the above calculations for generators of
rotations.

Let us now turn to the remaining part of the (deformed) Poincar\'e
symmetry, namely the symmetries with respect to the space time
translations. We have
$$
\varepsilon P_\mu\triangleright S=\int d^{4}x\,\int\int d^4
p\,e^{3p_0/\kappa}\,d^4 k \, e^{3k_0/\kappa }
$$
\begin{equation}\label{Q5}
    \varepsilon(p_\mu \kp k_\mu)\Phi(p)\Psi(k)\mathcal{M}(k):e^{ipx}::e^{ikx}:=0
\end{equation}
where we used definition of delta function. following the same
procedure we can show invariance of the scalar product (\ref{F3}).

In terms of the Fourier transformed fields the action reads
$$
 S = \frac12 \int\, d p_0 d^3 \mathbf{p}\,  {\Phi}(p_0, \mathbf{p})\, {\Phi}(-p_0, - e^{p_0/\kappa} \,\mathbf{p})\,
$$
\begin{equation}\label{29}
  \left[ \kappa^2\sinh^2\frac{p_0}{\kappa} -  \frac12\, \mathbf{p}^2\left( e^{2p_0/\kappa} +1\right) + \frac{\mathbf{p}^4}{4\kappa^2}\, e^{2p_0/\kappa} - M^2\right]
\end{equation}
where we used the explicit form of ${\cal M}(p_0,\mathbf{p})$
\begin{equation}\label{30}
 {\cal M}(p_0,\mathbf{p}) \equiv  \left[ \kappa^2\sinh^2\frac{p_0}{\kappa} -  \frac12\, \mathbf{p}^2\left( e^{2p_0/\kappa} +1\right) + \frac{\mathbf{p}^4}{4\kappa^2}\, e^{2p_0/\kappa} - M^2\right]
\end{equation}
 Note that  the factors $e^{3p_0/\kappa}$ in the integration measure
cancel. This action can be also expressed in slightly more compact
way using the antipode $S$ (generalized ``minus'') of the
$\kappa$-Poincar\'e algebra defined in equations (\ref{4})
\begin{equation}\label{29a}
  S = \frac12 \int\, d p_0 d^3 \mathbf{p}\,  {\Phi}(p_0, \mathbf{p})\, {\Phi}(S(p_0), S(\mathbf{p}))\, {\cal M}(p_0,\mathbf{p})
\end{equation}
Varying this action with respect to $\Phi(p_0,
\mathbf{p})$ and noting that
$$
{\cal M}(p_0,\mathbf{p})
= {\cal
M}(S(p_0),S(\mathbf{p}))
$$   we find the on shell condition of
the form
\begin{equation}\label{30a}
 \kappa^2\sinh^2\frac{p_0}{\kappa} -  \frac12\, \mathbf{p}^2\left( e^{2p_0/\kappa} +1\right) +
 \frac{\mathbf{p}^4}{4\kappa^2}\, e^{2p_0/\kappa} - M^2=0
\end{equation}
Now we can write down the real field on shell in the following form
\begin{equation}\label{c2}
    \Phi(x)=\int d^{3} k\,a^{*}_{k}e^{i\mathbf{k}\mathbf{x}}e^{-ik_0 x_0}+\int d^3 k\,a_{k}e^{iS(\mathbf{k})\mathbf{x}}e^{-iS(k_0) x_0}
\end{equation}
which is very similar to the classical case. The difference is the
antipode S in exponent instead of minus. On the mass shell we have
\begin{equation}\label{c4}
    e^{k_0/\kappa}=\frac{1}{1-\frac{|\mathbf{k}|}{\kappa}}.
\end{equation}
where for simplicity we consider only the massless case.

\section{conclusions}

We have shown
 that in analogy to the classical  case, for scalar
fields on $\kappa$-Minkowski one can construct scalar product and
Hermitian action of $\kappa$-Poinar\'e algebra generators. A
subalgebra of this algebra can be interpreted as infinitesimal
Lorentz transformations, under which the field action is
invariant. The solutions of $\kappa$-deformed Klein-Gordon
equation have the form similar to classical field.


\begin{thebibliography}{99}


\bibitem{zakrz} S.~Zakrzewski, ``Quantum Poincar\'e group related to
the $\kappa$-Poincar\'e algebra,'' J.\ Phys.\ {\bf A27}, 2075
(1994)

\bibitem{MajidPLB334} S.~Majid and H.~Ruegg, "Bicrossproduct structure of kappa Poincare group
and noncommutative geometry," Phys.\ Lett.\ B {\bf 334} (1994) 348

\bibitem{qP} J.~Lukierski, H.~Ruegg, A.~Nowicki and V.~N.~Tolstoi,
``Q deformation of Poincar\'e algebra,'' Phys.\ Lett.\ B {\bf 264}
(1991) 331; J.~Lukierski, A.~Nowicki and H.~Ruegg, ``New quantum
Poincare algebra and k deformed field theory,'' Phys.\ Lett.\ B
{\bf 293} (1992) 344.

\bibitem{Amelino-Camelia:2000ge}
G.~Amelino-Camelia, ``Testable scenario for relativity with minimum-length,''
Phys.\ Lett.\ B {\bf 510}, 255 (2001) [arXiv:hep-th/0012238].
\bibitem{jkgminl} J.~Kowalski-Glikman,
``Observer independent quantum of mass,'' Phys.\ Lett.\ A {\bf 286} (2001) 391
[arXiv:hep-th/0102098].

\bibitem{rbgacjkg} N.~R.~Bruno, G.~Amelino-Camelia and J.~Kowalski-Glikman,
``Deformed boost transformations that saturate at the Planck scale,'' Phys.\
Lett.\ B {\bf 522} (2001) 133 [arXiv:hep-th/0107039].

\bibitem{Kowalski-Glikman:2002jr}
J.~Kowalski-Glikman and S.~Nowak, ``Non-commutative space-time of doubly
special relativity theories,'' Int.\ J.\ Mod.\ Phys.\ D {\bf 12} (2003) 299
[arXiv:hep-th/0204245].

\bibitem{Kowalski-Glikman:2003hi}
J.~Kowalski-Glikman,
``Doubly special relativity and quantum gravity phenomenology,''
arXiv:hep-th/0312140, Proceedings of 10th Marcel Grossmann Meeting, to appear; J.~Kowalski-Glikman,
``Introduction to doubly special relativity,''
arXiv:hep-th/0405273, Lecture Notes in Physics, to appear.


\bibitem{Amelino-Camelia:2002vy}
G.~Amelino-Camelia, ``Doubly-Special Relativity: First Results and Key Open
Problems,'' Int.\ J.\ Mod.\ Phys.\ D {\bf 11} (2002) 1643
[arXiv:gr-qc/0210063];
G.~Amelino-Camelia,
 ``Some encouraging and some cautionary remarks on doubly special relativity in
quantum gravity,''
arXiv:gr-qc/0402092.

\bibitem{Matschull:1997du}
H.~J.~Matschull and M.~Welling, ``Quantum mechanics of a point
particle in 2+1 dimensional gravity,'' Class.\ Quant.\ Grav.\ {\bf
15} (1998) 2981 [arXiv:gr-qc/9708054].

\bibitem{Freidel:2003sp}
L.~Freidel, J.~Kowalski-Glikman and L.~Smolin,
``2+1 gravity and doubly special relativity,''
Phys.\ Rev.\ D {\bf 69} (2004) 044001
[arXiv:hep-th/0307085].

\bibitem{Kosinski:1998kw}
P.~Kosinski, P.~Maslanka, J.~Lukierski and A.~Sitarz,
Czech.\ J.\ Phys.\  {\bf 48} (1998) 1407.

\bibitem{Kosinski:1999ix}
P.~Kosinski, J.~Lukierski and P.~Maslanka,
``Local D = 4 field theory on kappa-deformed Minkowski space,''
Phys.\ Rev.\ D {\bf 62} (2000) 025004
[arXiv:hep-th/9902037].

\bibitem{Kosinski:2001ii}
P.~Kosinski, J.~Lukierski and P.~Maslanka,
 ``kappa-deformed Wigner construction of relativistic wave functions and  free
fields on kappa-Minkowski space,''
Nucl.\ Phys.\ Proc.\ Suppl.\  {\bf 102} (2001) 161
[arXiv:hep-th/0103127].

\bibitem{Amelino-Camelia:2001fd}
G.~Amelino-Camelia and M.~Arzano, ``Coproduct and star product in field
theories on Lie-algebra  non-commutative space-times,'' Phys.\ Rev.\ D {\bf 65}
(2002) 084044 [arXiv:hep-th/0105120]; G.~Amelino-Camelia, M.~Arzano and L.~Doplicher,
``Field theories on canonical and Lie-algebra noncommutative spacetimes,''
arXiv:hep-th/0205047.


\bibitem{Dimitrijevic:2004nv}
M.~Dimitrijevic, L.~Jonke, L.~Moller, E.~Tsouchnika, J.~Wess and M.~Wohlgenannt,
``Field theory on kappa-spacetime,''
arXiv:hep-th/0407187;
M.~Dimitrijevic, F.~Meyer, L.~Moller and J.~Wess,
``Gauge theories on the kappa-Minkowski spacetime,''
Eur.\ Phys.\ J.\ C {\bf 36} (2004) 117
[arXiv:hep-th/0310116];
M.~Dimitrijevic, L.~Jonke, L.~Moller, E.~Tsouchnika, J.~Wess and M.~Wohlgenannt,
``Deformed field theory on kappa-spacetime,''
Eur.\ Phys.\ J.\ C {\bf 31} (2003) 129
[arXiv:hep-th/0307149].

\bibitem{Kosinski:2003xx}
P.~Kosinski, P.~Maslanka, J.~Lukierski and A.~Sitarz,
``Generalized kappa-deformations and deformed relativistic scalar fields  on
noncommutative Minkowski space,''
arXiv:hep-th/0307038.


\bibitem{Daszkiewicz:2004xy}
M.~Daszkiewicz, K.~Imilkowska, J.~Kowalski-Glikman and S.~Nowak,
``Scalar field theory on kappa-Minkowski space-time and doubly
special relativity,'' arXiv:hep-th/0410058.







\bibitem{GAChopf}
A.~Agostini, G.~Amelino-Camelia and F.~D'Andrea, ``Hopf-algebra
description of noncommutative-spacetime symmetries,''
arXiv:hep-th/0306013.

\bibitem{Woron}
S.~L.~Woronowicz, "Differential Calculus on Compact Matrix Pseudogroups (Qauntum Groups),"
Commun.\ Math.\ Phys.\ {\bf 122}, 125-170 (1989).

\bibitem{Gon}
C.~Gonera and P.~Kosiñski, P.~Maœlanka, "Differential calculi on quantum Minkowski space,"
J.\ Math.\ Phys. {\bf 37} (11), November 1996.

\bibitem{Sitarz}
A.~Sitarz, "Noncommutative differential calculus on the $\kappa$-Minkowski space,"
Phys.\ Lett.\ B {\bf 349} (1995) 42-48.

\bibitem{deSitter}
J.~Kowalski-Glikman and S.~Nowak, "Doubly special relativity and de Sitter space,"
Class.\ Quant.\ Grav. {\bf 20} (2003) 4799 [arXiv:hep-th/0304101]

\end{thebibliography}
\end{document}